\documentclass{article}

\usepackage[final, nonatbib]{neurips_2022}
\usepackage[utf8]{inputenc} 
\usepackage[T1]{fontenc}    
\usepackage{hyperref}       
\usepackage{url}            
\usepackage{booktabs}       
\usepackage{amsfonts}       
\usepackage{nicefrac}       
\usepackage{microtype}      
\usepackage[table]{xcolor}

\usepackage{xcolor}         
\usepackage{graphicx}
\usepackage{multirow}
\usepackage{array}
\usepackage{amssymb}
\usepackage{amsmath}
\usepackage{caption}
\usepackage{bbding}
\usepackage{subcaption}
\usepackage{blindtext}
\usepackage{adjustbox}

\usepackage{amsmath}

\usepackage{graphicx}

\usepackage{amssymb}
\usepackage{pifont}

\newcolumntype{L}[1]{>{\raggedright\let\newline\\\arraybackslash\hspace{0pt}}m{#1}}
\newcolumntype{C}[1]{>{\centering\let\newline\\\arraybackslash\hspace{0pt}}m{#1}}
\newcolumntype{R}[1]{>{\raggedleft\let\newline\\\arraybackslash\hspace{0pt}}m{#1}}
\usepackage[utf8]{inputenc}
\usepackage[english]{babel}
\usepackage{amsthm}
\usepackage[misc]{ifsym}

\newcommand{\app}{\raise.17ex\hbox{$\scriptstyle\sim$}}
\makeatletter\renewcommand\paragraph{\@startsection{paragraph}{4}{\z@}
  {.495em \@plus1ex \@minus.2ex}{-.5em}{\normalfont\normalsize\bfseries}}\makeatother

\makeatletter
\DeclareRobustCommand\onedot{\futurelet\@let@token\@onedot}
\def\@onedot{\ifx\@let@token.\else.\null\fi\xspace}

\newcommand{\eg}{\mbox{e.g.,\ }}

\newcommand{\ie}{\mbox{i.e.,\ }}

\makeatother

\newcolumntype{x}[1]{>{\centering\arraybackslash}p{#1pt}}
\newcolumntype{a}[1]{>{\columncolor{verylightgray}\centering\arraybackslash}p{#1pt}}
\newcolumntype{y}[1]{>{\raggedright\arraybackslash}p{#1pt}}
\newcolumntype{z}[1]{>{\raggedleft\arraybackslash}p{#1pt}}\newlength\savewidth

\newcolumntype{P}[1]{>{\centering\arraybackslash}p{#1}}

\usepackage{bbding}
\usepackage{bibunits}
\usepackage{bm}
\usepackage{array}
\usepackage{xcolor}
\usepackage{pdflscape}
\usepackage{makecell}
\usepackage{framed,multirow}
\usepackage[flushleft]{threeparttable}
\usepackage{amssymb}
\usepackage{pifont}
\usepackage{algorithm}
\usepackage{layouts}
\usepackage{listings}
\usepackage{pythonhighlight}
\makeatletter
\newcommand\footnoteref[1]{\protected@xdef\@thefnmark{\ref{#1}}\@footnotemark}
\makeatother

\definecolor{Highlight}{HTML}{39b54a}  

\title{Synthetic Tumors Make AI Segment Tumors Better}

\author{
Qixin~Hu\textsuperscript{1}~~~~~~Junfei~Xiao\textsuperscript{2}~~~~~~Yixiong~Chen\textsuperscript{3}~~~~~~Shuwen~Sun\textsuperscript{4}~~~~~~Jie-Neng~Chen\textsuperscript{2}\\\textbf{Alan~Yuille}\textsuperscript{\textbf{2}}~~~~~~\textbf{Zongwei~Zhou}\textsuperscript{\textbf{2,}}\thanks{Corresponding author: Zongwei Zhou (\href{mailto:zzhou82@jh.edu}{zzhou82@jh.edu})} \\[2mm]
\textsuperscript{1}Huazhong University of Science and Technology \quad
\textsuperscript{2}Johns Hopkins University \\
\textsuperscript{3}Fudan University \quad \textsuperscript{4}The First Affiliated Hospital of Nanjing Medical University \\[2mm]
{\small Code and Visual Turing Test:~\href{https://github.com/MrGiovanni/SyntheticTumors}{https://github.com/MrGiovanni/SyntheticTumors}}
}

\begin{document}
\include{math_command}
\maketitle

\begin{abstract}
    We develop a novel strategy to generate synthetic tumors.
    Unlike existing works, the tumors generated by our strategy have two intriguing advantages:
    (1) realistic in shape and texture, which even medical professionals can confuse with real tumors;
    (2) effective for AI model training, which can perform liver tumor segmentation similarly to a model trained on real tumors---this result is \textit{unprecedented} because no existing work, using synthetic tumors only, has thus far reached a similar or even close performance to the model trained on real tumors.
    This result also implies that manual efforts for developing per-voxel annotation of tumors (which took years to create) can be considerably reduced for training AI models in the future.
    Moreover, our synthetic tumors have the potential to improve the success rate of small tumor detection by automatically generating enormous examples of small (or tiny) synthetic tumors.
    
\end{abstract}

\section{Introduction}
\label{sec:introduction}

Artificial intelligence (AI) has dominated medical image segmentation~\cite{zhou2018unet++,zhou2019unet++,isensee2021nnu}, but training an AI model (\eg U-Net~\cite{ronneberger2015u}) often requires a large number of detailed per-voxel annotations. Annotating medical images is not only expensive and time-consuming, but also requires extensive medical expertise, and sometimes needs the assistance of radiology reports and biopsy results to precisely annotate a tumor~\cite{zhou2021towards,wang2021development}. Due to its high annotation cost, only a total of roughly 100 CT scans with annotated liver tumors are publicly available (provided by LiTS~\cite{bilic2019liver}) for training and testing models.

\begin{figure*}[!t]
    \centering
    \includegraphics[width=1.0\linewidth]{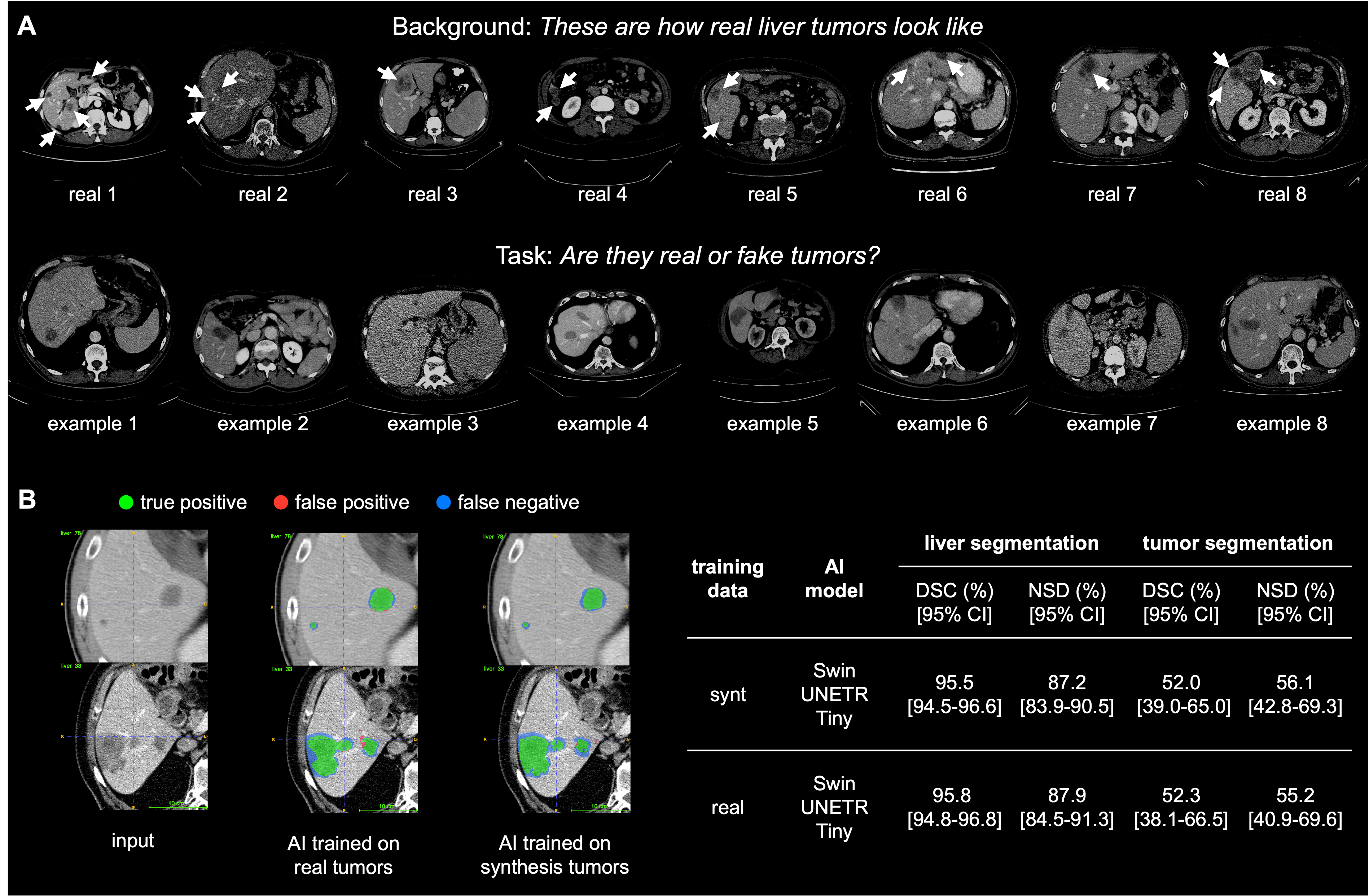}
    \caption{
    \textbf{A.} We conduct an examination for humans to distinguish synthetic tumors from the real ones (\ie Visual Turing Test). In these examples, some CT scans contain real liver tumors, and some contain synthetic tumors generated by our algorithm. \textbf{B.} Qualitative assessment and quantitative analysis of AI models trained on synthesis tumors and real tumors. Both of them can obtain similar segmentation performance while generating synthetic tumors requires no manual annotation cost.
    }
    \label{fig:main_results}
\end{figure*}

Generating synthetic tumors is an attractive research topic. There are some early attempts at generating COVID-19 infections on Chest CT scans~\cite{yao2021label}, abdominal tumors in CT scans~\cite{jin2021free}, diabetic lesions on retinal images~\cite{wang2022anomaly}, brain tumors on MRI images~\cite{wyatt2022anoddpm}, and cancers in fluorescence microscopy images~\cite{horvath2022metgan}. 
However, the synthetic tumors in those existing studies appear very different from the real tumors, and AI models trained using synthetic tumors perform significantly worse than those trained using real tumors due to the pronounced domain gap between real and synthetic tumors. 
\textit{What makes synthesizing tumors so hard?} 
There are several important factors: shape, intensity, size, location, and most importantly, texture.
In this paper, we develop a hand-crafted heuristic strategy to synthesize liver tumors in abdominal CT scans.
Our synthetic tumors are realistic---even medical professionals can confuse them with real tumors in the Visual Turing Test~\cite{geman2015visual,han2019synthesizing} (\figureautorefname~\ref{fig:main_results}A).
Besides, AI models trained on our synthetic tumors can segment real tumors similar to those trained on real tumors with expensive, detailed per-voxel annotation. 
As shown in \figureautorefname~\ref{fig:main_results}B, the model trained on our (label-free) synthetic tumors achieves a Dice Similarity Coefficient (DSC) of 52.0\% for segmenting real liver tumors, whereas AI trained on real tumors obtains a DSC of 52.3\% (no statistical difference between the two performances). These results are unprecedented because no existing work, using synthetic tumors \textit{only}, has thus far reached a similar performance to the model trained on real tumors.

\begin{figure*}[!t]
    \centering
    \includegraphics[width=1.0\linewidth]{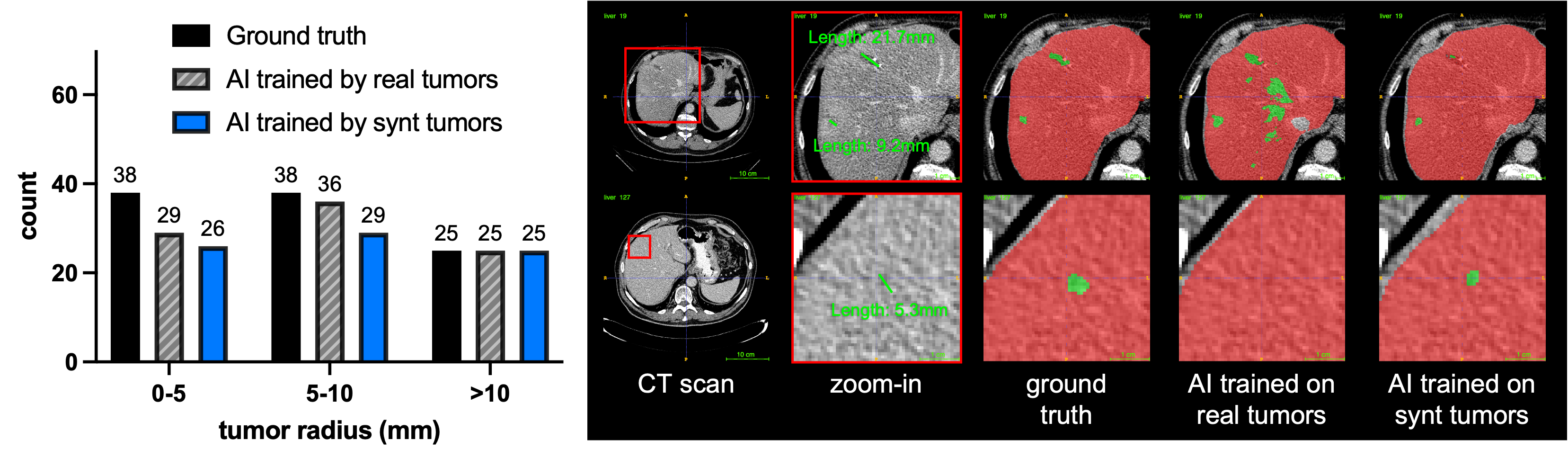}
    \caption{
    	Performance of liver tumor detection stratified by tumor size. The left panel presents the tumor-level sensitivity. For both models, the false negatives are mostly smaller than 10mm. The right panel presents two examples of small tumors. The smallest tumor we detected was 2mm. 
    }
    \label{fig:tumor_size_detection}
\end{figure*}

More importantly, synthesizing tumors also enables us to exhaustively generate tumors of desired locations, sizes, shapes, textures, and intensity, which are not limited to a fixed finite-size training set. For example, it is hard to gather sufficient training examples with annotated small tumors since they often occur at the early stage of cancer, and present subtle abnormal textures to human eyes, and therefore it is difficult for experts to specify the boundary of the tumor manually. In contrast, our synthesizing strategy can generate enormous examples with small (or tiny) tumors, so it can potentially detect small tumors more effectively (see~\figureautorefname~\ref{fig:tumor_size_detection}). In summary, our ultimate goal is ambitious: to train AI models for tumor segmentation without using any manual annotation---this study makes a significant step towards it.

\section{Materials \& Methods}
\label{sec:methodology}

\textbf{\textit{Dataset \& Metric.}}
Detailed per-voxel annotations of liver tumors are provided in the LiTS dataset~\cite{bilic2019liver}.
The tumor types include HCC and secondary liver tumors and metastasis derived from colorectal, breast, and lung cancer.
The volume of liver tumors ranges from 38mm$^3$ to 349 cm$^3$, and the radius of tumors is approximately in the range of [2, 44]mm.
We split LiTS into a training set (101 CT scans) and a test set (22 CT scans), which follows the conventional setting used in the literature (\eg \cite{tang2022self}).
An AI model (\ie Swin UNETR-Tiny\footnote{Swin UNETR is a hybrid segmentation architecture, which integrates the benefits of both U-Net~\cite{ronneberger2015u} and Transformer~\cite{dosovitskiy2020image,liu2021swin}. We base our experiments on Swin UNETR because it is very competitive and has ranked top one in numerous public benchmarks~\cite{tang2022self}, including liver tumor segmentation (MSD-Liver).}~\cite{hatamizadeh2022swin}) is trained on the 101 CT scans with annotated liver tumors.
For comparison, a dataset of 109 CT scans with a healthy liver is assembled from CHAOS~\cite{kavur2021chaos} (20 CT scans), BTCV~\cite{landman2015} (47 CT scans), and Pancreas-CT~\cite{TCIA_data} (42 CT scans). We then generate liver tumors in these scans on the fly, resulting in enormous image-label pairs of synthetic tumors and their masks for training the AI model.
To evaluate the model's segmentation performance, we calculate the Dice similarity coefficient (DSC) and Normalized Surface Dice (NSD) with 2mm tolerance to quantify the performance.

\textbf{\textit{Tumor Generator.}} We develop a hand-crafted heuristic strategy to generate synthetic liver tumors, consisting of shape generation and texture generation. First, ellipse generation, elastic deformation, and mask blurring are sequentially applied to generate the sphere-like shape of tumors. Second, salt noise, Gaussian filtering, scaling, and clipping are applied to generate tumor-like texture. This is the basic version of our tumor generator, wherein the hyper-parameters are adjusted by (1) visual inspection between the real and synthetic tumors, (2) standardized guidance of the Liver Imaging Reporting and Data System (LI-RADS)~\cite{m2021use}, and (3) feedback from clinicians during the early iterations of the Visual Turing Test (introduced below). We have implemented several critical feedback into our tumor generator, \eg the mass effect and cirrhosis on the healthy part of the liver around the synthetic tumor; the satellite effect of multiple small tumors around a large tumor.

\textbf{\textit{Visual Turing Test}} is adopted to assess the quality of synthetic tumors from professionals' perspectives (examples in Figure~\ref{fig:main_results}A). 
The generated synthetic tumors are considered realistic if the professionals fail to distinguish them from the real tumors. Our preliminary results show an accuracy of 60\% (30/50), performed by a professional with 6-year experience, but a more evaluation is required.

\section{Results \& Discussion}
\label{sec:results_discussion}

Both qualitative and quantitative results in~\figureautorefname~\ref{fig:main_results}B demonstrate that the model trained on synthetic tumors performs similarly to the model trained on real tumors in segmenting real liver tumors from unseen CT scans.
Specifically, the Swin UNETR-Tiny achieves DSC scores of 52.0\%~[95\% CI: 39.0\%-65.0\%] and 52.3\%~[95\% CI: 38.1\%-66.5\%] when trained on synthetic and real tumors, respectively. A slightly higher NSD score (56.1\%~vs.~55.2\%) achieved by the model trained on synthetic tumors indicates that the model can also detect the boundary of the liver tumor precisely. Moreover, we evaluate the performance of small tumor detection. \figureautorefname~\ref{fig:tumor_size_detection} stratifies tumors by different sizes and plots the detection rates of models trained on synthetic and real tumors. Both models are capable of detecting liver tumors that are larger than 10mm radius. As of now, the model trained on real tumors (65 out of 76 real tumors detected) outperforms the model trained on synthetic tumors (55 out of 76 detected) in detecting tumors smaller than 10mm.
We anticipate the ability of small tumor detection can be improved by generating more small-sized synthetic tumors in the training stage. This is one of the advantages of synthetic tumors because CT scans with small liver tumors are very difficult to collect and annotate in clinical practice.
The right panel of~\figureautorefname~\ref{fig:tumor_size_detection} presents two examples of small tumors that are successfully detected by the model, and visually, the segmentation quality is greater than the ground truth.

\textbf{\textit{Conclusion.}} 
In this paper, we proposed an unsupervised strategy to generate realistic shapes and textures of liver tumors. Synthetic tumors enable AI models to perform similarly to the model trained on real liver tumors---collecting and annotating real tumors can take years to complete, but our proposed strategy is label-free. This reveals the great potential for to use of synthesis tumors to train AI models on larger-scale healthy CT datasets (which are much easier to obtain than CT scans with liver tumors). In practice, we can generate enormous synthetic tumors in CT scans, which facilitate annotation-efficient AI development and allow us to assess AI's capability of detecting tumors of varying locations, sizes, shapes, intensities, and textures.

\clearpage

\smallskip\noindent\textbf{Acknowledgements.} This work was supported by the Lustgarten Foundation for Pancreatic Cancer Research.
We thank Camille Torrico and Alexa Delaney for improving the writing of this paper.

\bibliographystyle{plain}
{\footnotesize
  \bibliography{refs}
}

\end{document}